\documentclass[numbers=noenddot]{jfm}
\usepackage[latin9]{inputenc}

\usepackage{array, ragged2e}
\usepackage{float}
\usepackage{mathtools}
\usepackage{graphicx}
\usepackage{esint}
\usepackage{verbatim}
\usepackage{caption}
\usepackage{subcaption}
\renewcommand\[{\begin{equation}}
\renewcommand\]{\end{equation}}
\newcommand{\un}[1]{\,\mathrm{#1}}
\numberwithin{equation}{section}
\usepackage{amsmath} 
\newcommand{\beq}{\begin{equation}}
\newcommand{\eeq}{\end{equation}}
\newcommand{\lb}{\left(}
\newcommand{\rb}{\right)}
\usepackage{tikz}
\usepackage{pgfplots}
\usepackage{overpic}
\usepackage{hyperref}
\usepackage{todonotes}
\author{Graham P. Benham\aff{1}\corresp{\email{graham.benham@ucd.ie}}, Olivier Devauchelle\aff{2}, Stuart J. Thomson\aff{3}}
\affiliation{\aff{1} School of Mathematics and Statistics, University College Dublin, Dublin 4, Ireland
\aff{2}Universit\'{e} de Paris, Institut de Physique du Globe de Paris, CNRS, F-75005 Paris, France
\aff{3}School of Engineering Mathematics and Technology, University of Bristol, Ada Lovelace Building, University Walk, Bristol, BS8 1TW,  UK}

\begin{document}

\title{On wave-driven propulsion}

\maketitle

\abstract{
A theory is presented for wave-driven propulsion of floating bodies driven into oscillation at the fluid interface. By coupling the equations of motion of the body to a quasi-potential flow model of the fluid, we derive expressions for the drift speed and propulsive thrust of the body which in turn are shown to be consistent with global momentum conservation. We explore the efficacy of our model in describing the motion of \emph{SurferBot} [Rhee \emph{et al.}\, \emph{Bioinspir. Biomim.}\ \textbf{17} (5), 2022], demonstrating close agreement with the experimentally determined drift speed and oscillatory dynamics. The efficiency of wave-driven propulsion is then computed as a function of driving oscillation frequency and the forcing location, revealing optimal values for both of these parameters which await confirmation in experiments. A comparison to other modes of locomotion and applications of our model to competitive water-sports are discussed in conclusion.
 }

\section{Introduction}
\label{sec_scal}

From flagellar beating of spermatozoa to water-walking insects, from the wave-like lateral flexions of fish to flying, nature has evolved myriad strategies for locomotion and propulsion in fluidic environments \citep{hu2003hydrodynamics,bush2006walking,buhler2007impulsive,gaffney2011mammalian,eloy2012optimal,smits2019undulatory}. A perhaps lesser-known propulsion mechanism is enacted by honeybees (\emph{Apis mellifera}) trapped on the surface of water: in oscillating its wings, the bee generates a fore-aft asymmetric wavefield that contributes in part to forward motion \citep{roh2019honeybees}. Inspired by the striken honeybee, \citet{rhee2022surferbot} designed {\it SurferBot}: a centimeter-scale interfacial robot that achieves wave-driven locomotion speeds of around $\sim$1 cm/s by inducing asymmetric vibrations with a small eccentric motor. Wave-driven propulsion may also be realized by designing floating bodies with mass- and/or geometric-asymmetries that generate thrust and torques by exchanging momentum with waves generated by the object as it oscillates on the free surface of a globally vibrating fluid bath \citep{ho2021capillary,barotta2023bidirectional}. At much larger scales, \citet{benham2022gunwale} achieved wave-driven locomotion speeds of around $\sim$1 m/s by forcing a canoe into oscillations by jumping up and down on its gunwales, a technique known as {\it gunwale bobbing}. In this case a simple wave-equation model was derived by treating the canoe as an oscillating pressure source with a prescribed motion, ignoring the effects of dispersion and considering an idealised Gaussian canoe shape. Although outside the scope of the present work, other studies have investigated alternative routes to wave-driven propulsion including symmetry-breaking due to the presence of boundaries \citep{tarr2023fluid} and using acoustic radiation forces \citep{sabrina2018shape,roux2022self,martischang2023acoustic}. 

\begin{figure}
\centering
\begin{tikzpicture}[scale=0.75]
\node at (0,0)  {\includegraphics[height=0.4\textwidth]{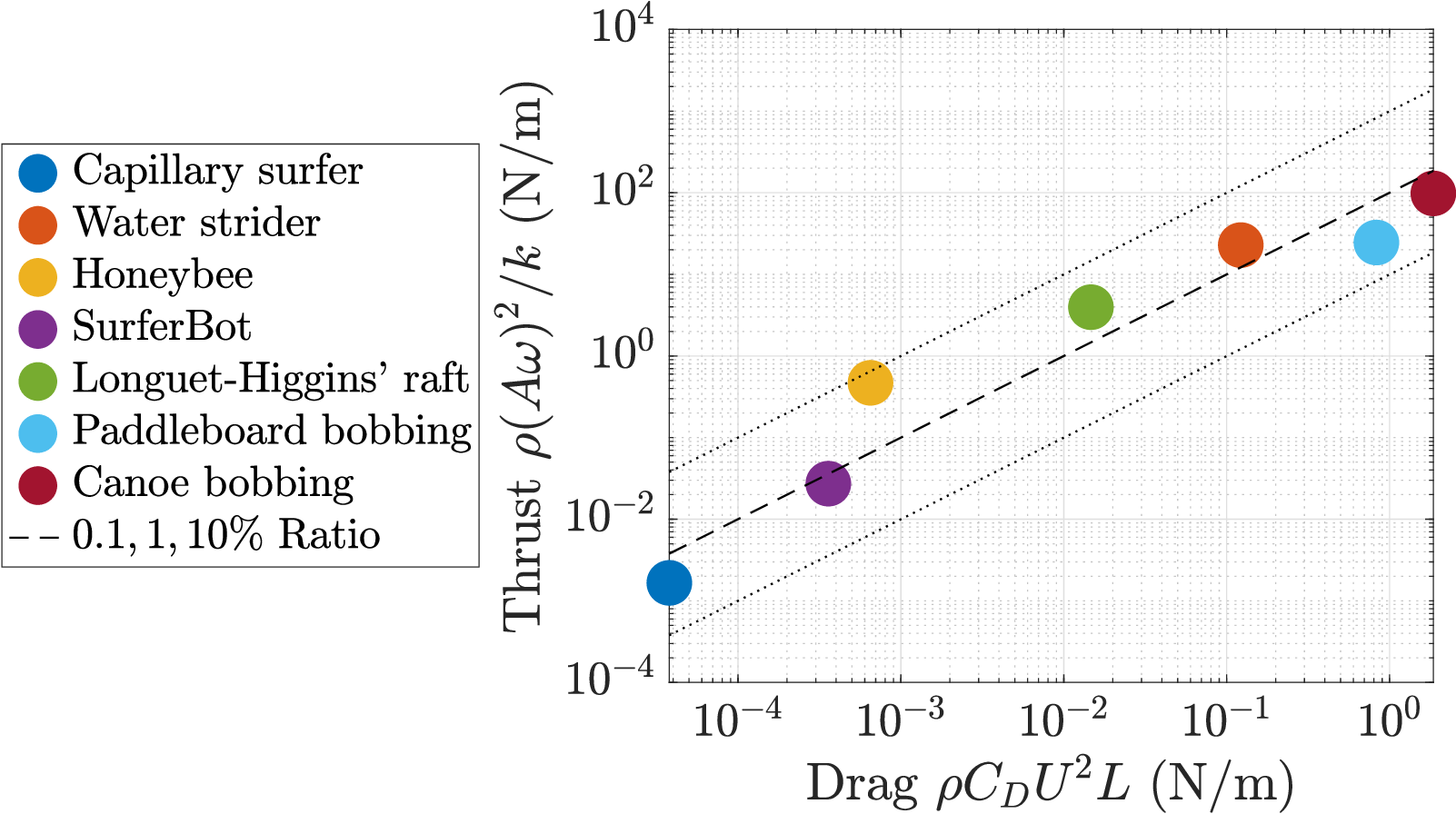}};
\end{tikzpicture}
\caption {Thrust scaling $F_T$ \eqref{scaleth} plotted against drag scaling $F_D$ \eqref{scaledr} for a variety of different bodies oscillating at the water surface (forces are taken as per unit width). Parameter values for each case are given in Table \ref{table1} in Appendix \ref{sec:table_data}. We observe that thrust balances drag with a ratio of $0.001-0.1$. 
\label{collaps} }
\end{figure}

Based on classical results \citep{longuet1964radiation,longuet1977mean}, the prevailing opinion is that the thrust necessary to drive locomotion originates from an asymmetric radiation of wave momentum \citep{ho2021capillary, rhee2022surferbot}. A scaling argument for two-dimensional wave-driven propulsion (i.e. restricted to the vertical plane) is as follows: Consider a body of length $L$ oscillating at the interface of a fluid with density $\rho$ and kinematic viscosity $\nu$ in the presence of gravity $g$. If the oscillations have amplitude $A$, frequency $\omega$, and wavenumber $k$, the thrust due to oscillating waves (per unit width) scales\footnote{This assumes that the fluid pressure applies over a distance $\sim 1/k$.} like
\beq
F_T\sim{\rho v^2/k},\label{scaleth}
\eeq
where $v=A\omega$ is the speed associated with the oscillations \citep{longuet1964radiation}. Forward motion at speed $U$ results in an inertial drag force that scales like
\beq
F_D\sim\rho C_DU^2L,\label{scaledr}
\eeq
where $C_D$ is a drag coefficient. By considering a number of known examples of wave-driven propulsion (Fig. \ref{collaps}), we demonstrate that thrust \eqref{scaleth} balances drag
\eqref{scaledr} with a ratio of $0.001-0.1$, indicating the range of prefactor values associated with \eqref{scaleth}.

Whilst the foregoing scaling argument performs well in capturing the thrust-drag balance, it fails to provide further insight into the problem. For example, there is no way of predicting the prefactors of the scalings or the related efficiency of propulsion. Further, without a model of the wavefield, momentum conservation cannot be verified. We herein present a theoretical model for wave-driven propulsion in the case of a raft floating at the fluid interface undergoing small-amplitude oscillations due to an external force. The equations of motion for the raft are coupled to a quasi-potential flow model of the fluid dynamics \citep{dias2008theory}. Propulsion is demonstrated in the form of a constant drift speed that results from a time-averaged thrust force due to the oscillations. We compare our predictions for the wavefield and raft motion with the experimental data of \emph{SurferBot} \citep{rhee2022surferbot}, and then use our model to optimize the efficiency of propulsion by varying the frequency of vibrational forcing and the position at which the forcing is applied. Some possible extensions and applications of our model to insect locomotion and competitive water sports are discussed in conclusion.

\section{Theoretical model for an oscillating raft}

Whilst scaling arguments for wave-driven propulsion show some success in predicting drag and thrust (see Fig.\ \ref{collaps}), there is no existing theory to model wave-driven dynamics that in turn can be used quantitatively to predict and maximise the efficiency of propulsion. In the following section we build such a model, first by formulating the equations of motion for a floating raft in two dimensions, then by accounting for the coupled fluid flow problem using quasi-potential flow theory. 

\begin{figure}
\centering
\begin{tikzpicture}[scale=0.65]
\node at (-10,0)  {\includegraphics[width=0.4\textwidth]{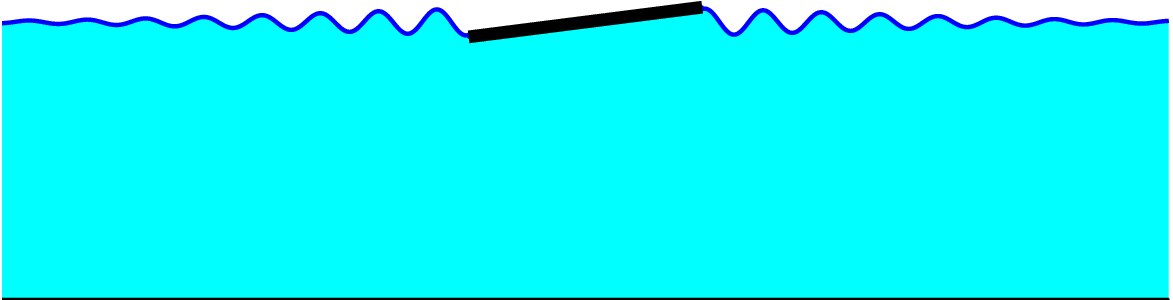}};
\node at (-12.85,1.3) {$z=\eta(x,t)$};
\node at (-12.95,-0.8) {$z=-H$};
\node at (-8.8,1.9) {$U$};
\draw[line width=1,black,->] (-9.85,1.5) -- (-7.85,1.5);
\draw[line width=1,dotted,black,->] (-5,0) -- (5,0);
\draw[line width=4,black] (-4,-1) -- (4,1);
\draw[line width=2,blue,<->] (0,-1) -- (0,1);
\draw[line width=2,red,<->] (-1,0) -- (1,0);
\draw[line width=2,<->,green] (4.1,0) .. controls (4.3,0.5) .. (4.1,1);
\node[red] at (1,-0.5) {$\xi(t)$};
\node[blue] at (0,1.3) {$\zeta(t)$};
\node[green] at (4.85,0.5) {$\theta(t)$};
\draw[line width=2,cyan,->] (3,0.8) -- (3,1.5) ;
\draw[line width=2,orange,->] (3,0.8) -- (3.8,0.8) ;
\node[cyan] at (2.6,1.85) {$F_{A,z}(t)$};
\node[orange] at (4.05,1.35) {$F_{A,x}(t)$};
\node at (3,0.8) {$\circ$};
\node at (1.5,0.9) {$x_A$};
\node at (-4.25,0.4) {$z=0$};
\node at (5.25,0) {$x$};
\draw[line width=1,<->] (0.05,0.2) -- (2.9,0.95);
\node at (-15,1.5) {(a)};
\node at (-5,1.5) {(b)};
\end{tikzpicture}
\caption{(a) A raft of length $L$ oscillating on the surface $z = \eta(x,t)$ of a fluid of depth $H$ self-propels with a drift-velocity $U$ thanks to a self-generated wavefield. (b) Schematic diagram of the raft dynamics (in the moving frame). The raft is subject to an oscillatory external force $\mathbf{F}_A = (F_{A,x}(t), F_{A, z}(t))$ applied at position $x_A$ (in the frame of the raft). The applied force gives rise to small horizontal and vertical oscillations $\xi(t)$ and $\zeta(t)$, in addition to planar rotations with angle $\theta$. \label{pitch} }
\end{figure}
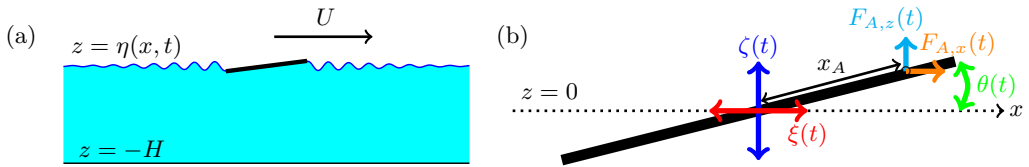

\subsection{Equations of motion for the raft}
\label{sec:EOM}
Let us consider a two-dimensional body of fluid with the horizontal and vertical coordinates given by $x$ and $z$, respectively. 
A floating raft of length $L$ resting on the fluid surface is subject to an applied periodic force $\mathbf{F}_A=(F_{A,x}(t),F_{A,z}(t))$ imposed at position ${\mathbf{x}}_A = (x_A, z_A)$ measured from the raft centre of mass (Figs. \ref{pitch}(a) and (b)). In response to $\mathbf{F}_A$ it is assumed that the centre of mass of the raft $\mathbf{X}$ undergoes small oscillations in the $x$- and $z$-directions, whilst translating horizontally with an \emph{a priori} unknown constant drift speed $U$. The action of $\mathbf{F}_A$ also induces a torque that gives rise to oscillations in the $x-z$ plane, the rotation vector given by $\boldsymbol{\theta}=\theta(t) {\mathbf{j}}$ where the unit vector ${\mathbf{j}}$ points out of the page. 

The equations of motion for the position and orientation of the centre of mass of the raft are
\begin{subequations}
\label{raft_dynamics}
\begin{align}
 m \ddot{\mathbf{X}}& = \mathbf{F}_A + \int_S  (p-p_a) \mathbf{n} \, \mathrm{d}x - \mathbf{F}_D,\label{zmomm}\\
 I \ddot{\boldsymbol{\theta}}& = {\mathbf{x}}_A\times \mathbf{F}_A  + \int_S  (p-p_a) \mathbf{x}\times \mathbf{n} \, \mathrm{d}x ,\label{thetamom}
\end{align}
\end{subequations}
where $m$ is the mass (per unit width) of the raft and $I=m L^2/12$ is moment of inertia of the raft (per unit width), taken as the value for a rod rotating about its centre of mass. The second term on the right-hand-side of Equation \eqref{zmomm} is the force arising from fluid pressure, $p$, minus its atmospheric value $p_a$, where the vector $\mathbf{n}$ is the unit normal to the raft surface, $S$, pointing in the positive $z$ direction. The drag force is approximated using an empirical drag coefficient such that
\beq
\mathbf{F}_D=\frac{1}{2}C_D \rho L |\dot{\mathbf{X}}|\dot{\mathbf{X}}.\label{dragterm}
\eeq
The torques on the right-hand-side of Equation \eqref{thetamom} are those arising from their corresponding forces in Equation \eqref{zmomm}. Drag on rotational motion is assumed to be negligible. We note the omission of a gravitational term in \eqref{zmomm} which is balanced by the pressure integral term in the steady state. While we do not go into details, the steady contribution to the pressure is a combination of surface tension (acting at the edges of the raft) and Archimedes buoyancy. We assume that the mass of the raft is such that this balance is sustained and hence we ignore the gravitational term from the momentum Equation \eqref{zmomm}. 

Consistent with experiments \citep{rhee2022surferbot}, the applied force is taken to be periodic and can be written as
\beq
(F_{A,x},F_{A,z})= (\hat{F}_{A,x},\hat{F}_{A,z}) e^{i\omega t}, 
\label{eqn:force_oscill}
\eeq
where $\hat{F}_{A,x}$, $\hat{F}_{A,z}$ are small complex constants (incorporating both amplitude and phase). In response, the position of the centre of mass of the raft is
\beq
\mathbf{X}=(X(t),Z(t))=(U t + \xi(t),\zeta(t)),
\eeq
where
\beq
(\xi,\zeta)=  (\hat{\xi},\hat{\zeta})e^{i\omega t},\qquad \hat{\xi},\hat{\zeta}\in\mathbb{C},
\label{eqn:oscillform}
\eeq
 represent small horizontal and vertical oscillations, respectively (Fig. \ref{pitch}(b)). The profile of the raft is thus described by
\beq
z=Z(t)+(x-X(t))\,\tan \theta(t)\quad\mathrm{on}:\quad |x-X(t)|\leq L/2,
\eeq 
where
\beq
\theta= \hat{\theta}e^{i\omega t},\qquad \hat{\theta}\in\mathbb{C}.
\label{eqn:oscillform2}
\eeq  
Likewise, the fluid pressure reads
\beq
p = p_a + \hat{p} e^{i\omega t},
\label{eqn:oscill_press}
\eeq
where $\hat{p} = \hat{p}(x, z)$ is a complex function.

Before going further, we pause to discuss the steady drift speed, $U$. The hypothesis of this paper is that steady drift is a result of a balance between the propulsion due to waves and the resultant inertial drag. We propose that the propulsion due to waves is given by the time-averaged $x$-component of the pressure integral term in \eqref{zmomm} and that this thrust force is given by
\beq
{\overline{F}}_{T}=\overline{\int_S(p-p_a)\lb \mathbf{n}\cdot \mathbf{i}\rb\,\mathrm{d}{x}}, \label{propest}
\eeq
where the overhead-bar notation denotes time-averaging and $\mathbf{i}$ is a unit vector in the horizontal. As shown in Appendix \ref{app_A}, the time-average of the drag term \eqref{dragterm} is well-approximated by the component due to steady drift $U$, while drag due to oscillations is comparatively small. Hence, the drag in the $x$-direction\footnote{We note that the time-averaged drag in the vertical direction is zero $\overline{\mathbf{F}}_{D}\cdot \mathbf{k}=0$, and so we write $\overline{\mathbf{F}}_{D}\cdot \mathbf{i}=\overline{F}_D$ to simplify notation.} approximates to
\beq
{\overline{F}}_{D}\approx \frac{1}{2}C_D \rho L U^2.\label{dragest0}
\eeq
Since the drag term \eqref{dragest0} is approximately the same as it would be for steady flow past the raft, we  use Blasius' theory of flow past a flat plate to estimate $C_D=1.33\cdot \mathrm{Re}^{-1/2}$ in terms of the Reynolds number Re $=UL/\nu$ \citep{schlichting2016boundary}. 

From the above analysis, it is clear that the time-averaged thrust \eqref{propest} and drag forces \eqref{dragest0} are both second order. In other words, if $\hat{F}_{A,x},\hat{F}_{A,z}=\mathcal{O}(\epsilon)$ then ${\overline{F}}_{D},{\overline{F}}_{T}=\mathcal{O}(\epsilon^2)$. Consequently, by taking a square root of the thrust-drag balance (${\overline{F}}_{T}={\overline{F}}_{D}$), we see that $U=\mathcal{O}(\epsilon)$.

To fully determine $U$ by equating \eqref{propest} and \eqref{dragest0} requires a fluid dynamics model for the pressure $p$ in Equation \eqref{propest}.
Likewise, the fluid pressure determines the first-order dynamics of the raft. Specifically, after substituting \eqref{eqn:force_oscill}, \eqref{eqn:oscillform}, \eqref{eqn:oscillform2} and \eqref{eqn:oscill_press} into Equations \eqref{raft_dynamics} and linearizing, the first-order oscillating dynamics
are described by
\begin{subequations}
\label{eqn:complex_all}
\begin{align}
 -m\omega^2 {\hat{\xi}}&= \hat{F}_{A,x},\label{complex1}\\
 -m\omega^2  {\hat{\zeta}}&=\hat{F}_{A,z} + \int_{-L/2}^{L/2} \hat{p}|_{{z}=0}  \, \mathrm{d}{x},\\
-\frac{1}{12}m\omega^2 \hat{\theta}&= {x}_A\hat{F}_{A,z}  + \int_{-L/2}^{L/2} {x}\hat{p}|_{{z}=0}  \, \mathrm{d}{x}.\label{complex3}
\end{align}
\end{subequations}
Equations \eqref{eqn:complex_all} provide a relationship between the constants $\hat{F}_{A,x},\hat{F}_{A,z}$, ${x}_A$, and the constants $\hat{\xi},\hat{\zeta},\hat{\theta}$, given $\hat{p}$ calculated by formulating the corresponding fluid flow problem.

\subsection{Quasi-potential flow}
The body of fluid is assumed to be irrotational and infinitely wide with a finite depth $H$ and a free surface at $z=\eta(x,t)$, such that $-\infty<x<\infty$ and $-H\leq z\leq\eta$. Following the approach of \citet{dias2008theory}, the velocity for a weakly damped\footnote{We choose to include viscosity in our model so as to replicate the spatial decay in waves observed in experiments (see Fig.\ \ref{wave_amp}).} flow can still be modelled as the gradient of a potential function $\phi(x,z,t)$ to good approximation. Due to incompressibility, $\phi$ satisfies
\beq
\nabla^2\phi=0.\label{lap_phi}
\eeq
The linearised kinematic condition applies on the water surface, such that 
\beq
\phi_z=\eta_t-2\nu \eta_{xx}:\quad\mathrm{on}\quad z=0.\label{bc1}
\eeq
On the raft the linearised kinematic condition takes the form
\beq
\phi_z =\dot{\zeta}+x\dot{\theta} :\quad\mathrm{on}\quad z=0,\quad |x-X|\leq L/2.\label{bc3}
\eeq
In addition, the unsteady version of Bernoulli's equation applies within the fluid which, upon linearisation, becomes
\beq
\phi_t + gz +\frac{p}{\rho} +2\nu \phi_{zz}=\frac{p_a}{\rho}.
\label{bc2old}
\eeq
After applying the dynamic boundary condition $p-p_a=-\gamma \eta_{xx}$ on the free surface, \eqref{bc2old} yields 
\beq
\phi_t + g\eta - \frac{\gamma}{\rho}\eta_{xx}+2\nu \phi_{zz}=0:\quad\mathrm{on}\quad z=0,\quad |x-X|> L/2,
\label{bc2}
\eeq
where Laplace pressure introduces the surface tension $\gamma$. Finally, the bottom surface is impermeable, such that
\beq
\phi_z= 0:\quad\mathrm{on}\quad z= -H. \label{bc4}
\eeq
In the same manner as \S\ref{sec:EOM}, we seek solutions of the form $\phi= \hat{\phi} e^{i\omega t}$, $
\eta= \hat{\eta} e^{i\omega t}$,
where $\hat{\phi} = \hat{\phi}(x,z)$, $\hat{\eta} = \hat{\eta}(x)$ are complex functions.

Henceforth, we formulate the fluid flow problem in the moving reference frame, $x\rightarrow x+Ut$. However, we restrict our attention to the case where the drift speed is much smaller than the typical wave speed, such that $U\ll\sqrt{gL}$. In this way, we focus on the flow resulting from the oscillations, rather than the drift velocity. 

The boundary conditions \eqref{bc1}, \eqref{bc3}, \eqref{bc2} and \eqref{bc4} are combined with \eqref{lap_phi} to give the governing system
\begin{subequations}\label{flow_oscill}
\begin{align}
{\nabla}^2\hat{\phi}&=0,\label{sys1}\\
\hat{\phi}_{z}&=\frac{\gamma}{\rho g} \hat{\phi}_{zxx}+\frac{\omega^2}{g}\hat{\phi}+4i \nu \omega  \hat{\phi}_{xx}  :\quad\mathrm{on}\quad z=0,\quad |x|> L/2,\\
\hat{\phi}_{z}&=i(\hat{\zeta}+x\hat{\theta}):\quad\mathrm{on}\quad z=0,\quad |x|\leq L/2,
\\
\hat{\phi}_{x}&= \pm i k\hat{\phi} :\quad\mathrm{on}\quad x= \mp \ell,  \label{sysrad}\\
\hat{\phi}_{z}&= 0:\quad\mathrm{on}\quad z= -H.\label{sysn}
\end{align}
\end{subequations}
For the purpose of numerical simulation, the domain is rendered finite of length $\ell$ in the $x$ direction. Radiative boundary conditions (Equation \eqref{sysrad}) are chosen at the left- and right-hand boundaries to avoid reflection, where $k$ satisfies the dispersion relation
\beq
\omega^2=k\tanh kH\lb g+\frac{\gamma}{\rho} k^2\rb +4i\nu \omega k^2.\label{disprelfinite}
\eeq
The final step in linking together the fluid-raft interaction is calculating the thrust force ${\overline{F}}_T$ \eqref{propest} in terms of the pressure and normal vector, noting that to leading order $\mathbf{n}\cdot \mathbf{i}\approx -{\eta}_{x}$.
From \eqref{bc2old} the pressure on the raft is given by
\beq
\hat{p} = -\rho(i\omega \hat{\phi} +g \hat{\eta}+ 2\nu \hat{\phi}_{zz})|_{z=0}.\label{pressure}
\eeq
Meanwhile, on the free surface the wavefield is given by the solution to the kinematic condition \eqref{bc1}, such that
\beq
i\omega\hat{\eta}-2\nu\hat{\eta}_{xx}=\hat{\phi}_{z}|_{z=0},\label{etafirst}
\eeq
which is accompanied by radiative boundary conditions (similar to \eqref{sysrad}) at $x=\mp \ell$ and continuity boundary conditions on the raft, namely $\hat{\eta}=\hat{\zeta}\pm\hat{\theta}L/2$ on $x=\pm L/2$. 

\subsection{Summary}
Let's now summarise the governing system of equations. Given constants $\hat{F}_{A,x}$,$\hat{F}_{A,z}$ and $x_A$, the dynamics of the raft are determined by Equations \eqref{eqn:complex_all}, coupled to the fluid flow problem given by Equations \eqref{flow_oscill} \emph{via} the pressure \eqref{pressure}. The drift speed $U$ is then calculated by equating thrust \eqref{propest} and drag \eqref{dragest0}, where ${\overline{F}}_T$ follows from the solution \emph{via} $\hat{p}$ and $\hat{\eta}$. The problem is solved numerically in MATLAB using a combination of finite differences\footnote{A large numerical domain is used, $\ell,H\gg L$, such that the solution does not depend on $\ell$ and $H$.} to solve the PDE system \eqref{flow_oscill} and Newton's method to resolve the coupling with \eqref{eqn:complex_all}. Example code is provided in the Supplemental Materials.

\section{Model results, validation and application to \emph{SurferBot}}

\begin{figure}
\centering
\begin{tikzpicture}[scale=0.75]
\node at (0,0)  {\includegraphics[height=0.23\textwidth]{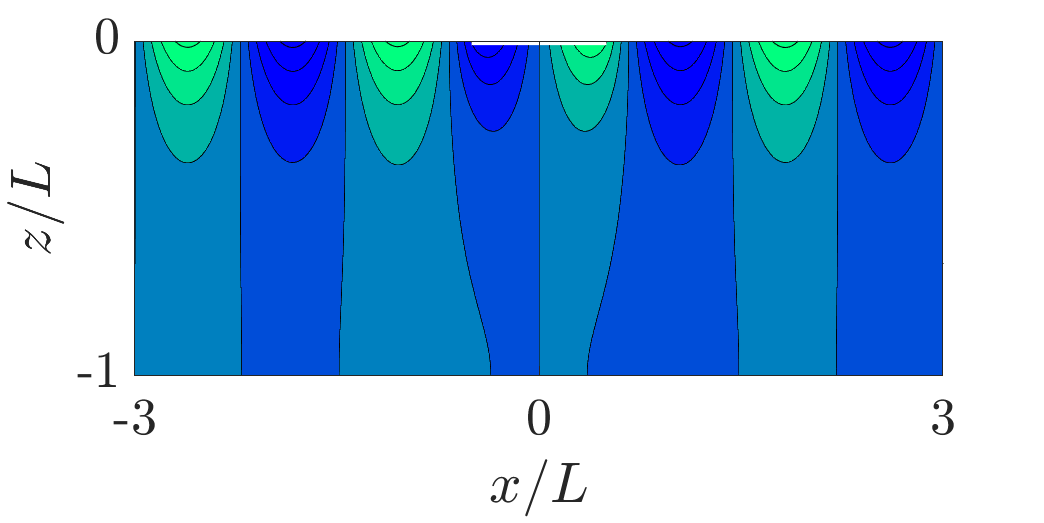}};
\node at (9,0)  {\includegraphics[height=0.23\textwidth]{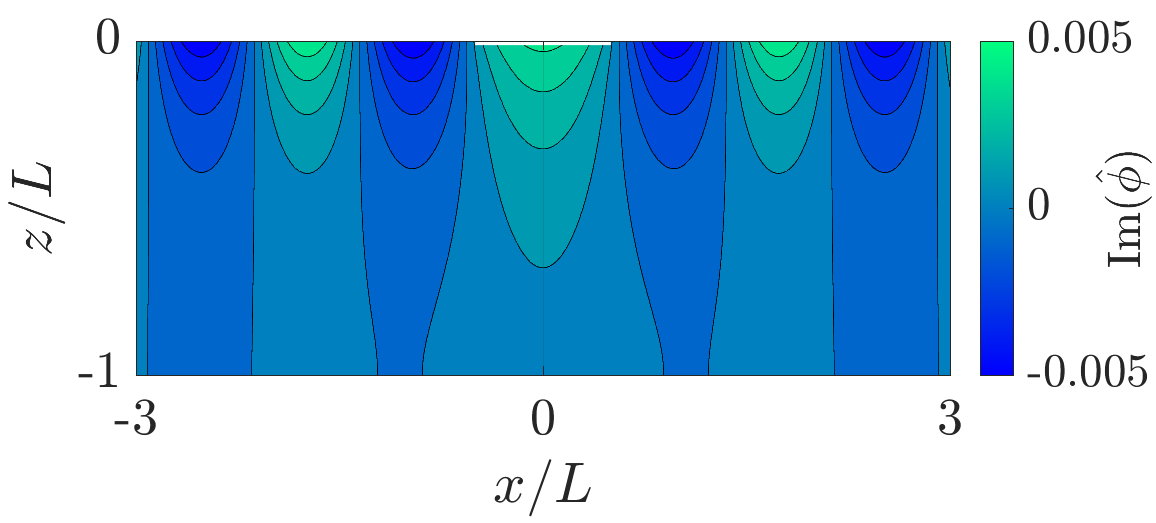}};
\node at (-3.75,2.2)  {(a)};
\node at (4.5,2.2)  {(b)};
\end{tikzpicture}
\caption{Velocity potential (taken as the imaginary part rescaled by $\omega L^2$) in the case of (a) pure pitching $\hat{\zeta}/L=0,\hat{\theta}=0.1,$ and (b) pure heaving $\hat{\zeta}/L=0.01,\hat{\theta}=0$, of a 1 m long raft oscillating at 1 Hz on the surface of water.
\label{phipic} }
\end{figure}

We begin this section by exploring the numerical results of our model, demonstrating that it is satisfies a conservation of momentum balance. We then consider a specific application of our model to \emph{SurferBot} \citep{rhee2022surferbot} comparing our results directly with experimental data.

\subsection{Numerical results and momentum balance}
Motivated by large-scale examples of bodies oscillating at the water surface \citep{benham2022gunwale}, as a simple demonstration of the fluid flow problem, we solve Equations \eqref{flow_oscill} numerically in the case of a 1 m long raft oscillating at 1 Hz on the surface of water ($\rho=1000\un{kg/m^3}$, $\gamma=0.073\un{N/m}$, $\nu=10^{-6}\un{m^2/s}$). For the sake of simplicity we restrict our attention to pure pitching ($\hat{\zeta}=0$) and pure heaving ($\hat{\theta}=0$), ignoring in the first instance the coupling to the forcing constants in Equations \eqref{eqn:complex_all}. The velocity potential is plotted in Fig. \ref{phipic} in the case of pure pitching (a) and pure heaving (b). As anticipated, $\overline{F}_T=0$ for both these cases since both heaving and pitching are required to generate thrust\footnote{Although both heaving and pitching are required to generate thrust in the case of vanishingly small drift velocity, it is not clear if this remains true at large drift velocities in general \citep{benham2022gunwale}. }.

\begin{figure}
\centering
\begin{tikzpicture}[scale=1]
\node at (0,0)  {\includegraphics[width=0.6\textwidth]{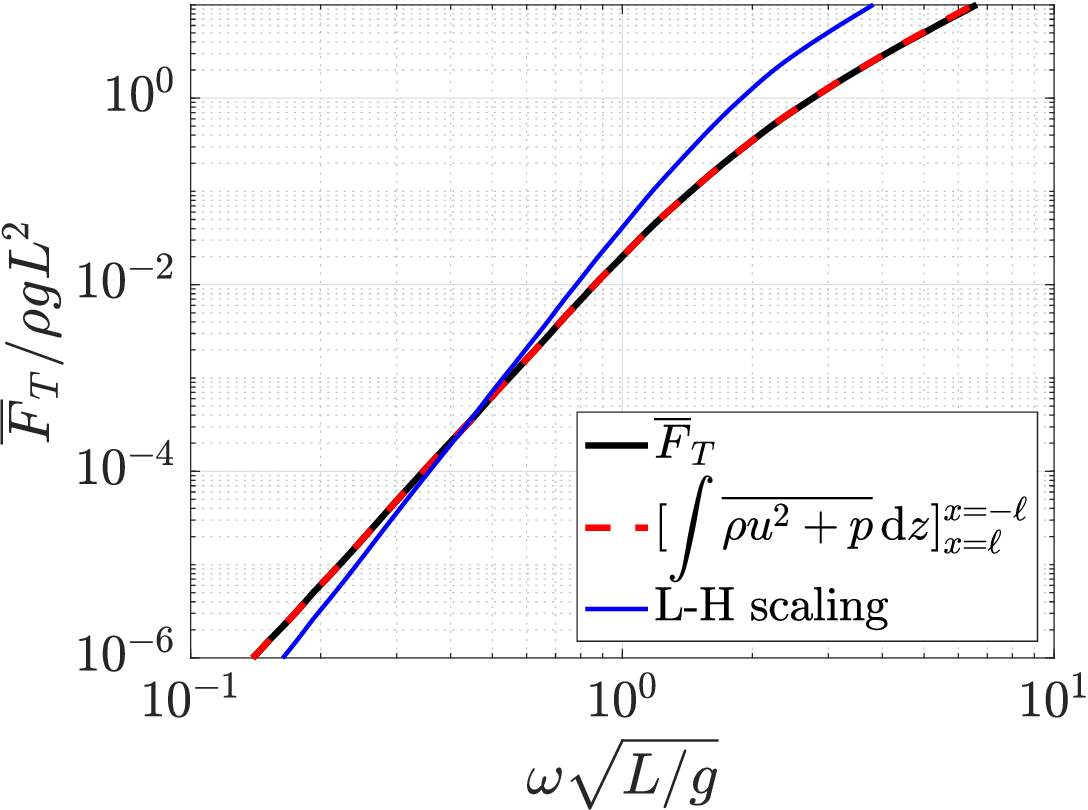}};
\end{tikzpicture}
\caption{Comparison between the numerically calculated thrust ${\overline{F}}_T$ and momentum flux across the domain (Equation \eqref{pretime}) showing close agreement between the two. Comparison is also made with the scaling proposed by Longuet-Higgins (see Appendix \ref{appC}) exhibiting only qualitative agreement. Surface tension and viscosity are neglected for the purpose of these calculations.
\label{mombal_app} }
\end{figure}

Next, we validate our numerical scheme using a conservation of momentum balance. As shown in Appendix \ref{appC} and first discussed by  \citet{longuet1964radiation}, an expression for the thrust force can be derived \emph{via} integration of the Euler equations. This results in an equivalence between $\overline{F}_T$ and the momentum flux of the form
\beq
{\overline{F}}_T=\left[\overline{\int_{-H}^{\eta}\left (\rho u^2 + p\right)\,\mathrm{d}z}\right]^{x=-\ell}_{x=\ell}.\label{pretime}
\eeq
The term on the right-hand side is the difference in mean momentum-flux between backward and forward directions (including the pressure difference). 
Hence, when larger waves propagate backwards than forwards, this results in a positive thrust. 
In Appendix \ref{appC} we show how to calculate this balance in the case where surface tension and viscosity are neglected. Calculating the left- and right-hand sides of Equation \eqref{pretime} numerically for a range of $\omega$ yields very close agreement between the two (Fig.\ \ref{mombal_app}), giving us confidence that the numerical scheme is accurate and consistent with global momentum conservation. 

It should be noted that whilst our model accurately conserves momentum, it neglects any possible contribution to thrust from vortex shedding since the fluid is considered irrotational. It is well-known that some forms of insect locomotion rely on both vortex shedding {\it and} capillary wave propagation \citep{bush2006walking, buhler2007impulsive}. To incorporate thrust from any vortices generated by the motion would require a model for rotational fluid flow, which we leave for a future study. However, by comparing our model with \emph{SurferBot} in the next section, we will see that our predictions perform well despite neglecting vorticity, suggesting that in the present case vortices are not the leading-order driver of propulsion \citep{barotta2023bidirectional}.

\subsection{Comparison with SurferBot}

\begin{figure}
\centering
\begin{tikzpicture}[scale=0.5]
\node at (-6.5,1.2) {\includegraphics[width=0.45\textwidth]{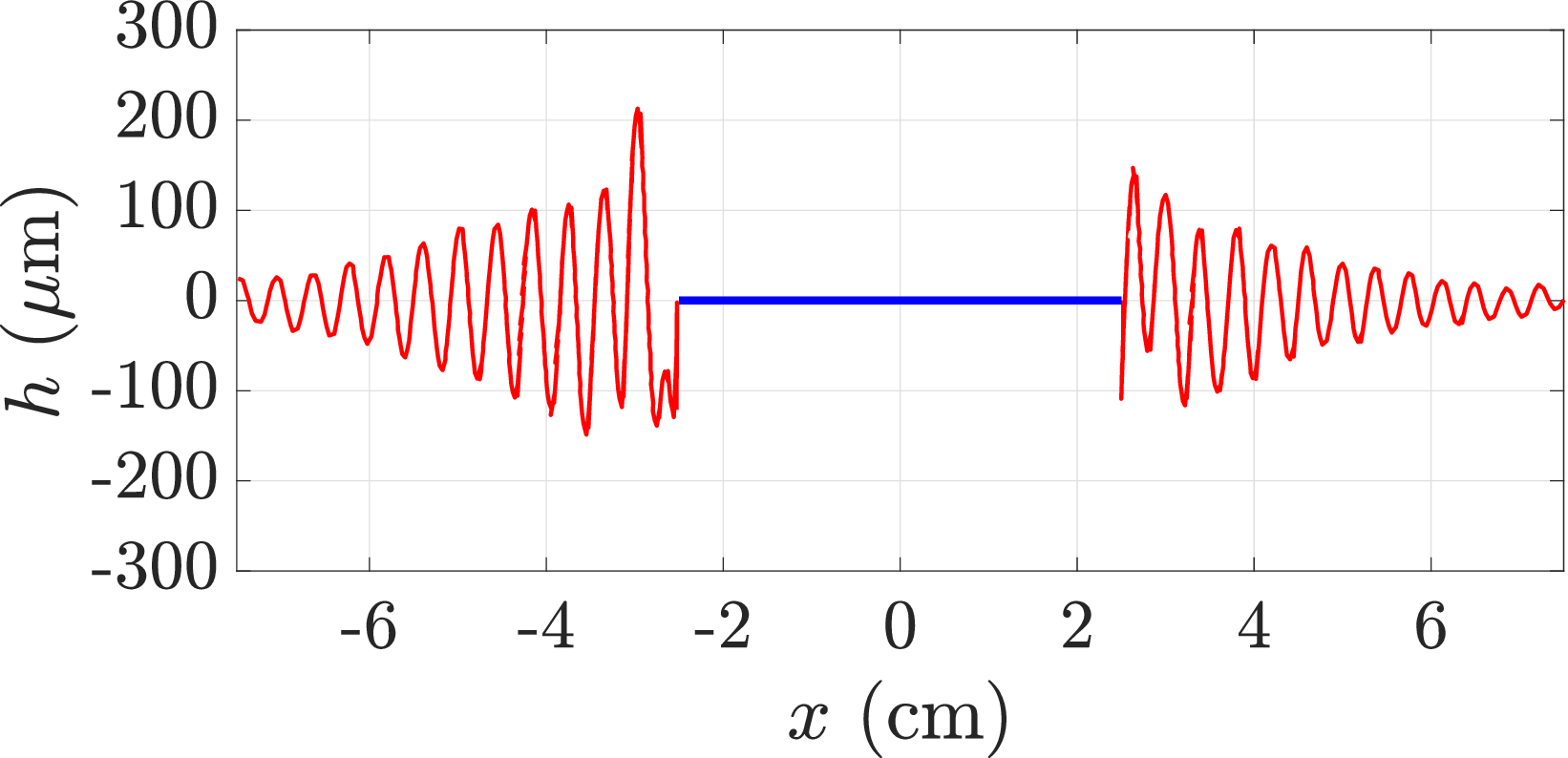}};
\node at (6.5,1.2) {\includegraphics[width=0.45\textwidth]{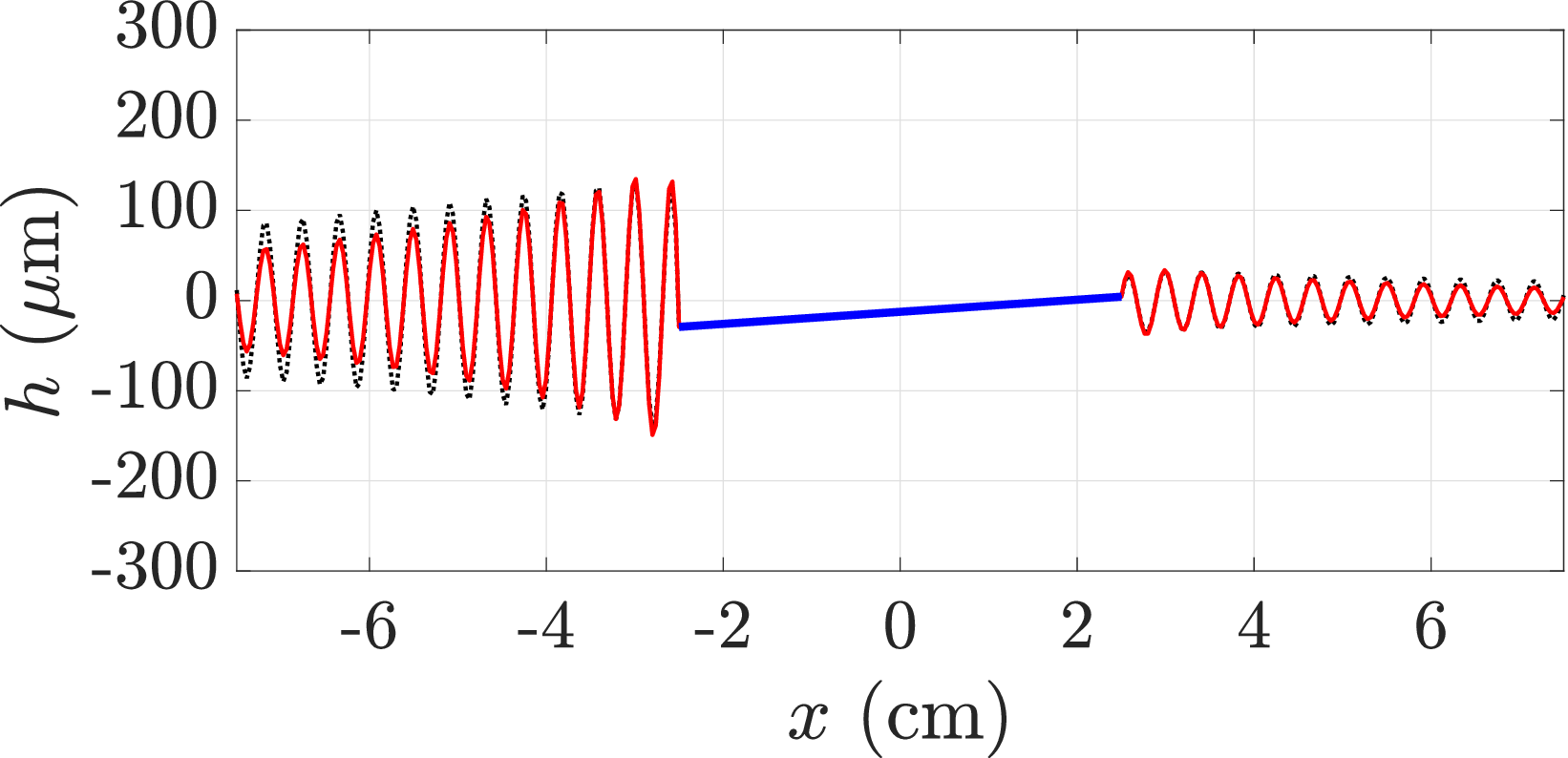}};
\node at (-6,4.5) {\large \it SurferBot};
\node at (7,4.5) {\large Present theory};
\node at (-12.5,3.7) {(a)};
\node at (0.5,3.7) {(b)};
\draw[line width=1,->,black] (-4,2.95) -- (-1,2.95);
\node at (-2.5,3.5) {\large $U$};
\node at (-6.5,-4.3) {\includegraphics[width=0.4\textwidth]{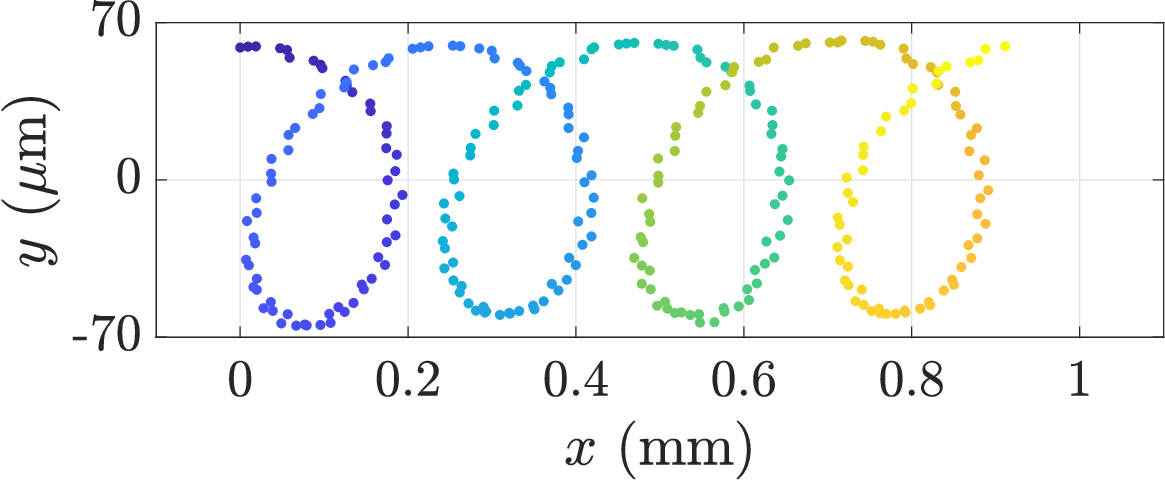}};
\node at (6.5,-4.3) {\includegraphics[width=0.4\textwidth]{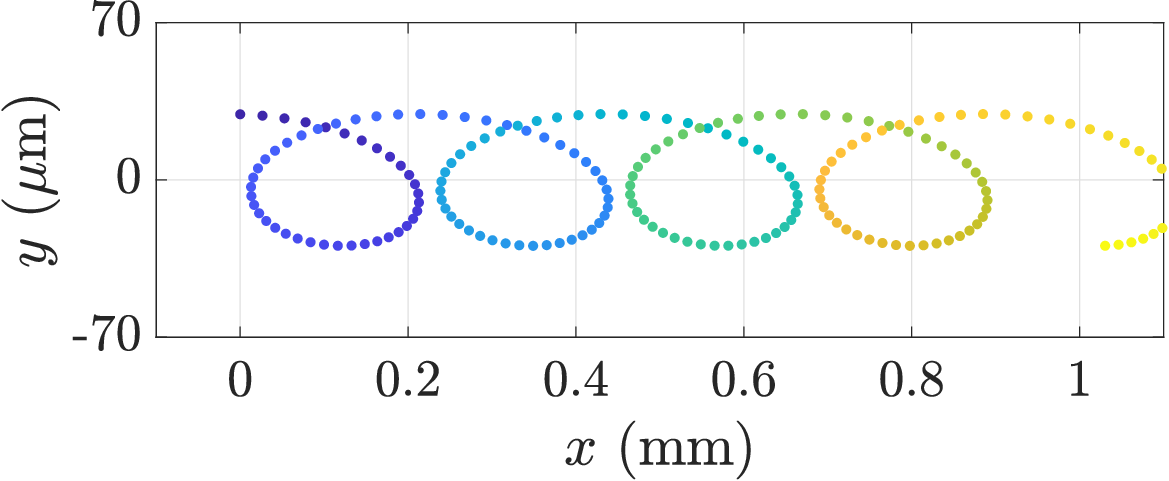}};
\node at (-6.5,-9) {\includegraphics[width=0.4\textwidth]{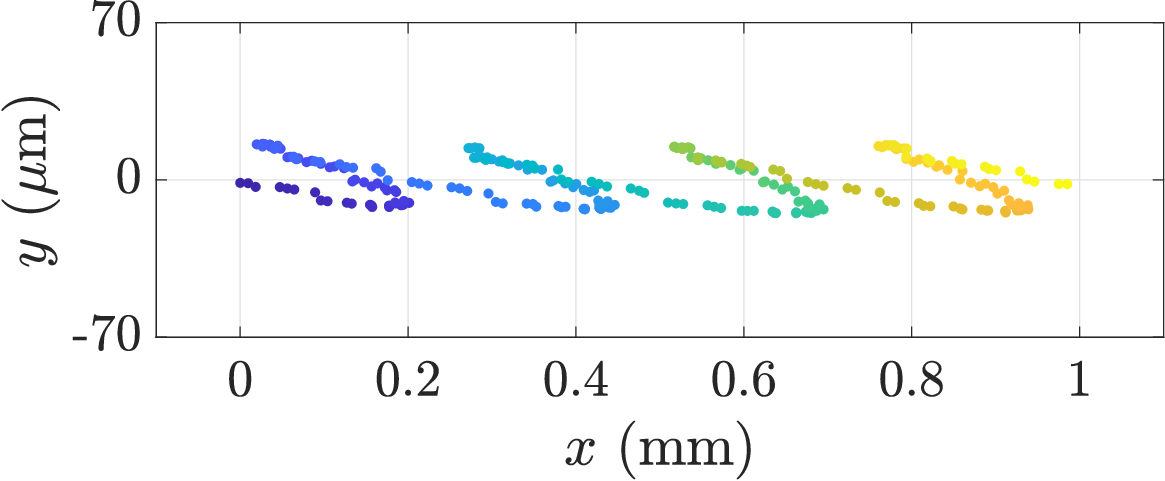}};
\node at (6.5,-9) {\includegraphics[width=0.4\textwidth]{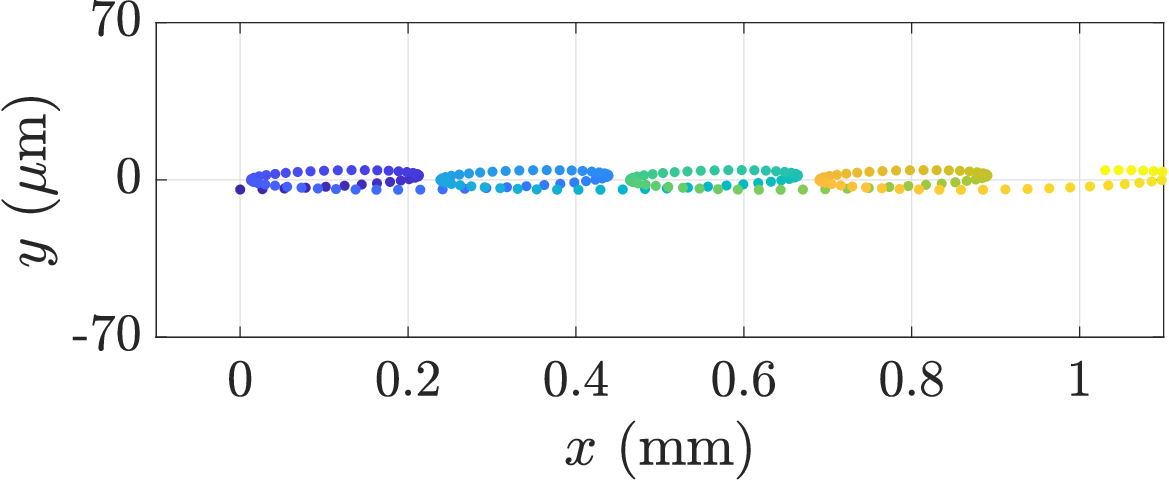}};
\node at (-12.5,-3.8) {\large \rotatebox{90}{$\mathbf{x}_\mathrm{aft}$}};
\node at (-12.5,-8.5) {\large \rotatebox{90}{$\mathbf{x}_\mathrm{fore}$}};
\node at (-12,-2) {(c)};
\node at (1.,-2) {(d)};
\node at (-12,-7) {(e)};
\node at (1.,-7) {(f)};
\node at (13.,-6) {\includegraphics[width=0.07\textwidth]{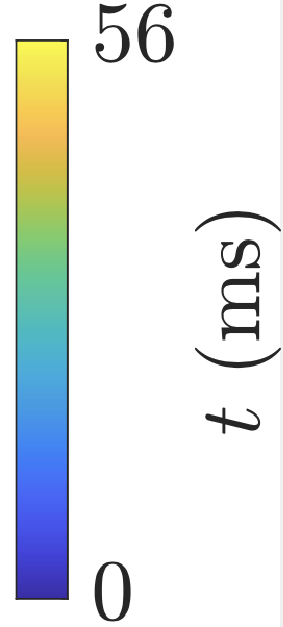}};
\end{tikzpicture}
\caption{Comparison between the experimental wave amplitude from (a) \emph{SurferBot} and (b) the mathematical model. In (b), the theoretical wave amplitude (red) is scaled by $\sqrt{2L/\pi x}$ to be consistent with the far-field behaviour of Bessel functions (since in practice the waves extend radially to the far-field). The unscaled wave amplitude is shown as a dotted black line. (c,d,e,f) Comparison between the experimental and theoretically predicted position of the aft (c,d) and fore (e,f). In the theoretical predictions (d) and (f), $\mathbf{x}_\mathrm{aft}=\mathbf{X}-(1/2,\theta/2)L$ and $\mathbf{x}_\mathrm{fore}=\mathbf{X}+(1/2,\theta/2)L$. Experimental data reproduced with permission. \label{wave_amp} }
\end{figure}
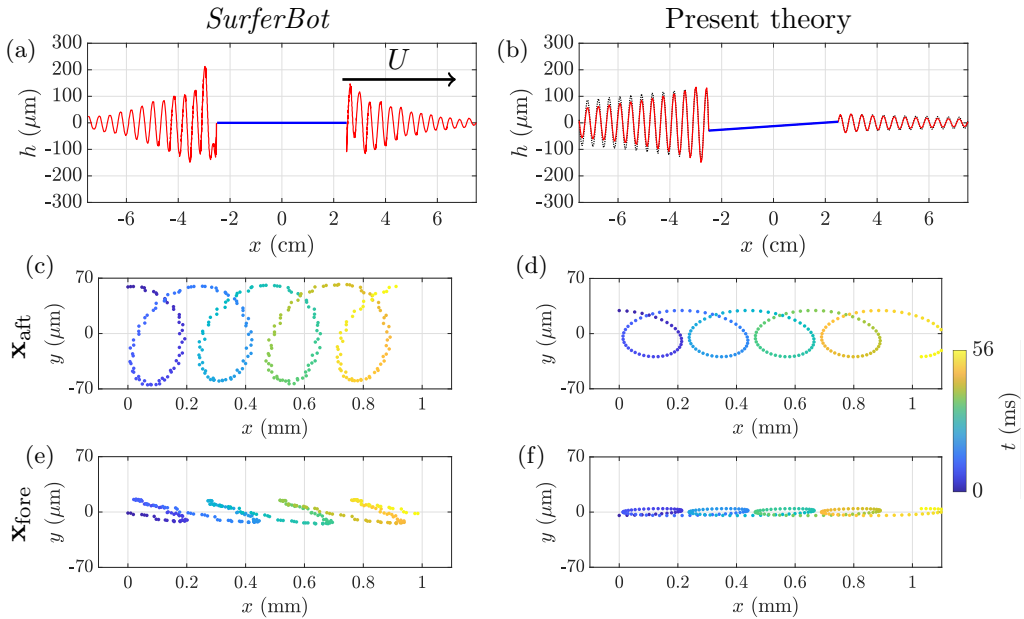

To demonstrate a specific application of our model, we calculate the dynamics of \textit{SurferBot} \citep{rhee2022surferbot}, for which dimensional parameter values are given in Table \ref{table1} in Appendix \ref{sec:table_data}. 
In this case, the force is applied \emph{via} an eccentric motor positioned at $x_A=-3\un{mm}$ from the centre of the raft. The forces $F_{A,x}$, $F_{A,z}$, are assumed to be out of phase, and are estimated using a scaling law relating the mass (per unit width) of the raft $m=2.6\un{g}/3\un{cm}$ and the oscillation amplitude $A=150\un{\mu m}$, such that
\beq
(\hat{F}_{A,x},\hat{F}_{A,z}) = m\omega^2A\,(i,1).\label{appforc}
\eeq
This results in a drift speed of $U\approx 2\un{cm/s}$, compared with an experimentally measured speed of $\sim1.8\un{cm/s}$. 
A comparison between the experimental and theoretical self-generated wavefield is plotted in Figures \ref{wave_amp}(a) and (b), whilst the time-varying position of the back and front of \emph{SurferBot} are shown in Figures \ref{wave_amp}(c,e) (experimental) and \ref{wave_amp}(d,f) (theoretical). Overall, the results of our theoretical model exhibit good qualitative agreement with experiments and yields an excellent prediction of the drift speed, $U$. However, there is disagreement in the amplitude of the wavefield and hence the oscillation amplitude of \emph{SurferBot}, in addition to the decay length-scale of the waves. However, these discrepancies are likely attributed to some obvious differences between experiment and theory, including the precise form of the oscillatory forcing, three-dimensional effects of both the \emph{SurferBot} geometry and wavefield, or indeed the flexure of the \emph{SurferBot} body contributing to wave damping.

\section{Efficiency and optimization}

Furnished with a theoretical model describing the \emph{SurferBot} dynamics, we can interrogate the efficiency of propulsion and optimize the parameters of \emph{SurferBot} thereafter. The efficiency of propulsion is given in terms of the applied and useful (propulsive) power. The total applied power is given by
\beq
{\overline{P}}_{A}= \overline{{F}_{A,z}\dot{\zeta}}+x_A \overline{F_{A,z}\dot{\theta}} , \label{ptot}
\eeq
the sum of linear and angular contributions\footnote{There is also a contribution $\overline{{F}_{A,x}\dot{\xi}}$ to ${\overline{P}}_{A}$ which vanishes because ${F}_{A,x}$ and $\dot{\xi}$ are out of phase due to \eqref{complex1}.}. 
The propulsive power is taken as $\overline{P}_T= \overline{F}_T U$ and the efficiency is defined as $\chi=\overline{P}_T/\overline{P}_{A}$. Hence, in the case of \emph{SurferBot} (before optimization), we calculate an efficiency value of $\chi=1.8\%$ from our mathematical model.

\begin{figure}
\centering
\begin{tikzpicture}[scale=1]
\node at (6.5,0) {\includegraphics[width=0.45\textwidth]{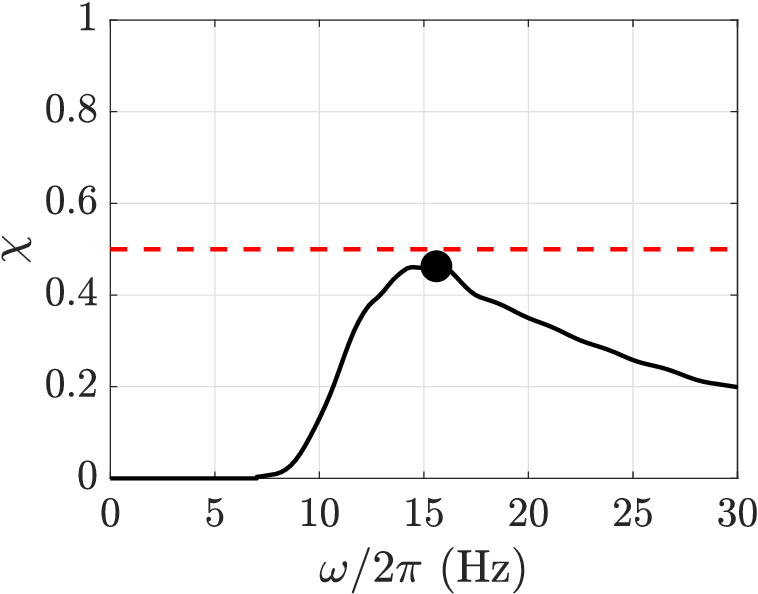}};
\node at (13,0) {\includegraphics[width=0.45\textwidth]{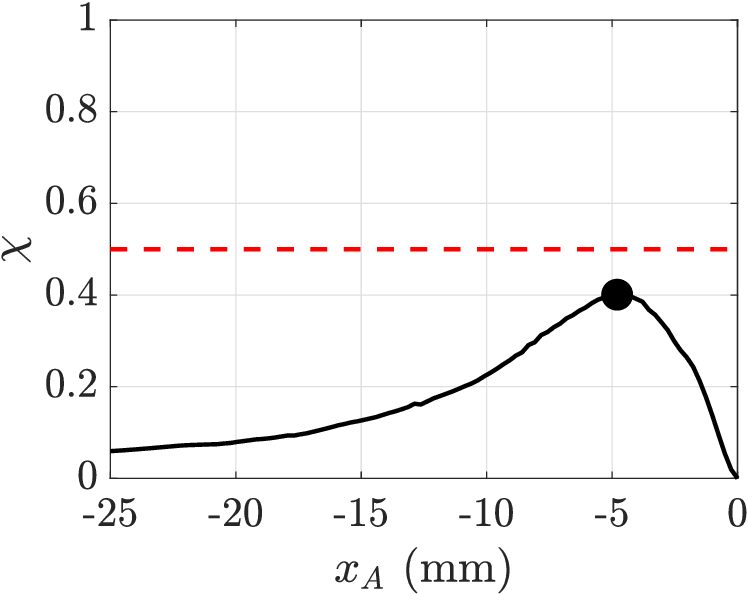}};
\node at (3.5,2.7) {(a)};
\node at (10,2.7) {(b)};
\node at (5.5,0.6) { \color{red} Subsonic limit};
\end{tikzpicture}
\caption{(a,b) Optimization of \emph{SurferBot} efficiency $\chi$ by varying the frequency and the motor position whilst maintaining constant total applied power $\overline{P}_{A}$. The optimum values (illustrated with black dots) are the outcome of a two-parameter optimisation, and hence the plotted curves are slices of the two-dimensional efficiency surface at its maximum.
\label{mombal} }
\end{figure}
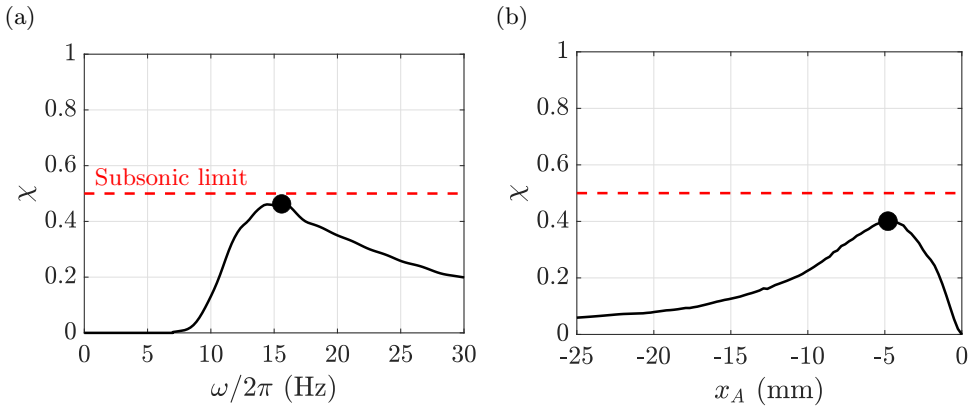

The two parameters we attempt to optimize are the driving frequency $\omega/2\pi$ and the horizontal position of the forcing, $x_A$. We assume that the horizontal force $\hat{F}_{A,x}=0$ (as it doesn't contribute to forward thrust anyway) and that the total power \eqref{ptot} is fixed\footnote{The applied force $\hat{F}_{A,z}$ must change to maintain constant power at variable frequency.} at the value calculated for \emph{SurferBot}, $\overline{P}_{A}=7.2\times 10^{-4} \un{W/m}$.  Efficiency is plotted as a function of $\omega/2\pi$ and $x_A$ in Figure \ref{mombal} showing an optimum frequency of 16 Hz and an optimum motor position of 5 mm behind the raft centre, both of which could be readily tested in a future experiment.

In a similar manner to the discussion in \S\ref{sec_scal}, it is also possible to derive a very rough estimate of efficiency using scaling arguments. To do so we may assume that the total power $\overline{P}_A$ \eqref{ptot} can be split into the contribution from propulsive power, which scales like $\overline{P}_T\sim \overline{F}_T U$,
and the power radiated away by the waves, which scales like $\overline{P}_{\mathrm{wave}}\sim \overline{F}_T c_g$, where $c_g=\mathrm{d}\omega/\mathrm{d}k$ is the group velocity of the waves. 
The efficiency then scales like
\beq
\chi\sim \frac{1}{1+1/\mathrm{Ma}},\label{chi_wave}
\eeq
where $\mathrm{Ma}=U/c_g$ is the Mach number. 
In the case of \emph{SurferBot} we find $\chi\sim 3.5\%$, which is in broad agreement with the model calculation $\chi=1.8\%$. 
However, such a scaling argument obviously provides no means for optimization.
Nevertheless, it indicates that efficiency may increase with Mach number, which is probably due to a decrease in forward-propagating waves as the motion approaches the supersonic limit (which we discuss further shortly). 
We also note that due to the scaling \eqref{chi_wave}, it is estimated that subsonic flows $\mathrm{Ma}<1$ have a maximum possible efficiency of $\chi \sim 1/2$, which we indicate with dashed lines in Figure \ref{mombal}.

\section{Discussion and perspectives}

Using linear quasi-potential flow theory, coupled with the equations of motion for an oscillating raft, we have demonstrated how the raft propels itself forward using its own waves. The close agreement between our linear model and the experimental data of \emph{SurferBot} \citep{rhee2022surferbot} indicates that we have captured the key physics at play for this self-propelling centimetre-scale robot. Nevertheless, there are several further improvements to the model that are worth discussing.

The assumption of a vanishingly small drift velocity is worth revisiting. In a future study a finite drift-speed could be incorporated into the fluid-flow problem. This would require transforming Equations \eqref{flow_oscill} to a moving reference frame, thereby introducing new derivative terms of the potential. Consequently, the dispersion relation \eqref{disprelfinite} would need to be modified and may result in multiple-wavenumber solutions for forward and backward travelling waves. An initial resolution to this problem could involve studying a small but finite drift speed using the method of asymptotic expansions. 

A finite drift-speed incorporated into the model would result in a Doppler-shifted wavefield, as discussed by \citet{benham2022gunwale}. In this case, a key dimensionless parameter is the Mach number, Ma, which is defined as the ratio between the drift speed $U$ and the group velocity of the waves $c_g$. The Mach number is closely related to whether or not \textit{forward} (in addition to backward) waves are propagated. Specifically, if the drift speed is larger than the wave speed for a particular wavenumber, this indicates that such a wave can only propagate behind the object and not in front of it.  
Since the momentum balance \eqref{pretime} indicates that larger forward waves reduce the forward thrust, we would expect more efficient motion at higher Mach numbers (which is consistent with the scaling \eqref{chi_wave}). The value of the Mach number and hence the shape of the wavefield varies considerably for different cases of wave-driven propulsion. For example, the wavefield of \emph{SurferBot} ($\text{Ma}=0.04$) is markedly different to that of gunwale bobbing on a paddleboard ($\text{Ma}=1.7$)\footnote{The Mach number is defined slightly differently from \citet{benham2022gunwale}, who used the phase velocity $c_p$ rather than the group velocity $c_g$ in the denominator, resulting in different quoted values.}. The former produces significant waves both forwards and backwards, whereas the latter has purely backward waves (e.g. compare Figure 3(a) of \citep{rhee2022surferbot} with Figure 1(a) of \citep{benham2022gunwale}).

By considering a finite drift-speed, one can then study how symmetry is broken when oscillations commence (i.e. the start-up problem for wave-driven propulsion). One could investigate, for example, how perturbations grow or decay from a stationary state when subjected to oscillations. This would then help determine the necessary conditions for initiating forward propulsion under different operating conditions. Therein lie numerous optimization problems, such as the optimal control of a single oscillator (or multiple oscillators) with variable positions or masses, not to mention the optimization of the shape of the raft itself. One could also consider different optimization objectives, such as maximising speed as well as efficiency.

A rotational fluid model could be developed to account for propulsion due to both vortex shedding and wave propagation. Whilst the present irrotational model performs well in the case of \emph{SurferBot}, it is well-known that vortices play a significant role in the locomotion of water striders which are smaller and faster than \emph{SurferBot} \citep{buhler2007impulsive}. At larger scales (e.g. in the case of gunwale bobbing) it is not known what role vorticity plays. However, by developing a rotational fluid flow model, one could establish the regimes in which vorticity can be neglected as a means of propulsion. 

One could extend the model to account for a three-dimensional flow field rather than the two spatial dimensions studied here. In such a model a more realistic representation of the \emph{SurferBot}  geometry (or other examples) could be rendered. This could then capture the anisotropic wavefield emitted by the rectangular-shaped \emph{SurferBot} and what effect this has on the thrust calculations studied here. Further, the model described herein assumes that the raft is subject to an external, \emph{local} force, motivated in part by the physical design of the \emph{SurferBot} and the eccentric mass motor that drives propulsion. An interesting extension to the present work would be to consider another form of symmetry-breaking wherein a raft with variable mass density along its length self-propels due to the \emph{global} vibration of the fluid bath \citep{ho2021capillary,barotta2023bidirectional}, a scenario akin to walking droplets \citep{couder2005walking, bush2015pilot}.

In closing, it is worth discussing some of the potential applications of this work. 
Firstly, as indicated by \citet{rhee2022surferbot}, studies on wave-driven propulsion may serve as a way to better understand how insects move when floating on the water surface. 
Secondly, boat oscillations during rowing and canoe races (due to the strokes of athletes) may affect performance due to wave-body interactions. For example, \citet{dode2022wave} studied how fluctuations in the horizontal speed may affect rowing efficiency. However, in the case of the vertical component of these oscillations, the impact on efficiency is uncertain \citep{benham2022gunwale}. Hence incorporating a finite drift-speed into the current model could be leveraged to optimize stroke styles to improve race times. 
Another extension of this work could include the interactions between multiple bodies. This may find application in the context of ducklings benefiting from their mother's wavefield \citep{yuan2021wave}, or in rowing/canoe races where boats may gain an advantage by interacting with the wavefields of their opponents.

\section*{Acknowledgements}
The authors would like to thank Jerome A.\ Neufeld for several helpful discussions in the early stages of the project, and Robert Hunt for providing the experimental \emph{SurferBot} data used in Figure \ref{wave_amp}.

\vspace{1cm}

The code used to solve this problem is found at: \\
\url{https://maths.ucd.ie/~gbenham/waves_code.zip}
\appendix
\section{Table of propulsion data}
\label{sec:table_data}
The data used in Figure \ref{collaps} of the main text is listed in Table \ref{table1}. To estimate the drag coefficients for these different cases we have used Blasius' theory of flow past a flat plate, $C_D=1.33\cdot \mathrm{Re}^{-1/2}$ in terms of the Reynolds number Re $=UL/\nu$ \citep{schlichting2016boundary}. This of course neglects the effects of the shape and aspect ratio, but nonetheless provides a useful scaling that applies across many orders of magnitude in length scale. 
The wavenumber $k$ is found for each case by solving the dispersion relation for deep water waves, $\omega^2=g k+\gamma k^3/\rho$.

\begin{table}
\centering
\begin{tabular}{|c|cccc|}
\hline
Type of motion  & $U$ (m/s) & $L$ (m) & $f$ (Hz) & $A$ (m) \\
  \hline
Capillary surfer \citep{ho2021capillary}
& 2.8$\times 10^{-3}$
& 6.4$\times 10^{-3}$
& 1$\times 10^{2}$
& 8.2$\times 10^{-5}$ \\
Water strider \citep{hu2003hydrodynamics}
& 1.5
& 1$\times 10^{-2}$
& 1$\times 10^{2}$
& 1$\times 10^{-2}$ \\
Honeybee \citep{roh2019honeybees}
& 4$\times 10^{-2}$
& 1.5$\times 10^{-2}$
& 60
& 2$\times 10^{-3}$\\
\emph{Surferbot} \citep{rhee2022surferbot}
& 1.8$\times 10^{-2}$
& 5$\times 10^{-2}$
& 80
& 150$\times 10^{-6}$ \\
Longuet-Higgins' raft \citep{longuet1977mean} 
& 0.125
& 0.25
& 3
& 2$\times 10^{-2}$ \\
Paddleboard bobbing  \citep{benham2022gunwale}
& 0.8
& 3.05
& 1.67
& 5$\times 10^{-2}$ \\
Canoe bobbing \citep{benham2022gunwale}
& 1.25
& 4
& 1.25
& 0.1 \\
\hline
\end{tabular}
\caption{Parameters used in Figure \ref{collaps} and their references. The type of fluid in all cases is water, for which
$\rho=1000\un{kg/m^3}$, $\gamma=0.073\un{N/m}$, $\nu=10^{-6}\un{m^2/s}$, except the capillary surfer, for which the fluid is a water-glycerol mixture and the parameters are $\rho=1176\un{kg/m^3}$, $\gamma=0.0664\un{N/m}$ and $\nu=1.675\times 10^{-5}\un{m^2/s}$. The frequency $f=\omega/2\pi$. \label{table1}}
\end{table}

\section{Approximation of drag term}
\label{app_A}
Drag is treated by inserting $\dot{X}=(U+\dot{\xi},\dot{\zeta})$ into Equation \eqref{dragterm} and taking the time average. In this way, we arrive at
\beq
\overline{\mathbf{F}}_D= \frac{1}{2}\rho C_D L^3\omega^2 \overline{\left|\begin{array}{c}
\frac{U}{\omega L}+i\frac{\hat{\xi}}{L}e^{i\omega t}\\
i\frac{\hat{\zeta}}{L}e^{i\omega t}
\end{array}\right|\lb\begin{array}{c}
\frac{U}{\omega L}+i\frac{\hat{\xi}}{L}e^{i\omega t}\\
i\frac{\hat{\zeta}}{L}e^{i\omega t}
\end{array}\rb}.
\eeq
We note that both $\hat{\xi}/L$ and $\hat{\zeta}/L$ are $\mathcal{O}(\epsilon)$. Hence, the drag in the $x$-direction can be Taylor expanded to give
\beq
\overline{\mathbf{F}}_D\cdot \mathbf{i}\approx \frac{1}{2}  C_D \rho L U^2
\eeq
which makes use of the fact that
\beq
\left(\frac{\omega L^2 }{U\hat{\zeta}}\right)^2\ll 1\label{asymcheck}
\eeq
is a small quantity (e.g. $(\omega L^2 /U\hat{\zeta})^2\approx 0.06$ in the case of \emph{SurferBot}).

\section{Momentum balance}
\label{appC}

We start by neglecting viscosity and writing down the two-dimensional (nonlinear) Euler equations for incompressible fluid flow in the presence of gravity, which are
\begin{align}
u_x+w_z&=0,\label{eul1}\\
u_t+uu_x+wu_z&=-\frac{1}{\rho}p_x,\label{eul2}\\
w_t + uw_x+ww_z&=-\frac{1}{\rho}p_z-g\label{eul3}.
\end{align}
By adding together \eqref{eul1} (multiplied by $u$) and \eqref{eul2} we get
\beq
u_t+\left[u^2\right]_x+\left[uw\right]_z=-\frac{1}{\rho}p_x.
\eeq
This can be integrated vertically to give
\beq
\int_{-H}^{\eta}\rho u_t \,\mathrm{d}z+\int_{-H}^{\eta}\left [\rho u^2 + p\right]_x\,\mathrm{d}z+\left.\rho u(\eta_t+u \eta_x)\right|_{z=\eta}=0,
\eeq
where we have used the impermeability and kinematic boundary conditions
\begin{align}
w&=0:\quad z=-H\\
w&=\eta_t+u \eta_x:\quad z=\eta(x,t).\label{kineq}
\end{align}
Next, we use the Leibniz rule and cancel terms to get
\beq
\frac{\partial}{\partial t}\int_{-H}^{\eta}\rho u \,\mathrm{d}z+\frac{\partial}{\partial x}\int_{-H}^{\eta}\left (\rho u^2 + p\right)\,\mathrm{d}z=\left. p \eta_x\right|_{z=\eta}.\label{pretime0}
\eeq
Let's now restrict our attention to oscillating waves which are periodic in time. Hence, by taking the time-average of \eqref{pretime0}, the first term vanishes. 
Meanwhile, the rest of the equation becomes
\beq
\frac{\mathrm{d}}{\mathrm{d} x}\left[\overline{\int_{-H}^{\eta}\left (\rho u^2 + p\right)\,\mathrm{d}z}\right]=\overline{(\left. p \eta_x\right|_{z=\eta})}.\label{b4xint}
\eeq
Next, we integrate from right to left (i.e. from $x=\ell$ to $x=-\ell$), giving 
\beq
\left[\overline{\int_{-H}^{\eta}\left (\rho u^2 + p\right)\,\mathrm{d}z}\right]^{x=-\ell}_{x=\ell}=\overline{F}_T,\label{pretime2}
\eeq
where $\overline{F}_T$ is the mean horizontal force generated by surface gradients
\beq
\overline{F}_T=\int_{-\ell}^\ell \overline{(-p \eta_x|_{z=\eta})} \,\mathrm{d}x.\label{app_F}
\eeq
This equates to Equationn \eqref{propest} upon linearisation of the normal vector. 

Next, we apply the foregoing calculation to a potential flow example in the absence of viscosity and surface tension, $\gamma=\nu=0$, since the calculation is much simpler. In the case of an oscillating raft, as described in the main text, the pressure within the fluid is given by the (linearised) Bernoulli's equation
\beq
p=p_a-\rho (g z + \phi_t).
\eeq
However, on the free surface $z=\eta$ (outside the raft region), the pressure is set by the dynamic boundary condition $p=p_a$. 
Hence, the force \eqref{app_F} becomes
\beq
\overline{F}_T= \rho \int_{-L/2 }^{L/2} \overline{(g \eta + \phi_t)\eta_x|_{z=0}} \,\mathrm{d}x.
\eeq
The term on the left hand side of \eqref{pretime2} approximates to
\beq
\left[\overline{\int_{-H}^{\eta}\left (\rho u^2 + p\right)\,\mathrm{d}z}\right]^{x=-\ell}_{x=\ell}\approx \left[\rho \int_{-H}^{0} \overline{\phi_x^2} \,\mathrm{d}z  \right]^{x=-\ell}_{x=\ell}.
\eeq
Hence, in the linear case the momentum balance \eqref{pretime2} takes the form
\beq
\left[ \int_{-H}^{0} \overline{{\phi}_{x}^2} \,\mathrm{d}z \right]^{x=-\ell}_{x=\ell} = \int_{-L/2 }^{L/2 } \overline{(g{\eta} + i\omega{\phi}){\eta}_{x}|_{z=0}} \,\mathrm{d}x.\label{linbal}
\eeq
We confirm this balance by numerical calculation in Figure \ref{mombal_app} of the main text, which shows extremely close agreement between the right and left hand sides of \eqref{linbal}. 
These calculations are for a fixed raft amplitude $A/L=1$ (which is achieved by setting $\hat{\zeta}/L=-1/2$, $\hat{\theta}=1$), whilst varying the frequency. All force values are normalised by the scaling $\rho g L^2$ for convenience. 

We also compare the numerically calculated thrust values with the scaling \eqref{scaleth}, incorporating the difference in amplitude between aft and fore waves. This provides the force scaling
\beq
{\overline{F}}_T\sim \frac{3\rho \omega^2[A^2]^{x=-\ell}_{x=\ell}}{4k},\label{LHscal}
\eeq
where we use a prefactor of 3/4 to be consistent with other authors \citep{ho2021capillary}. 
Since this scaling is attributed to Longuet-Higgins, we compare it to the numerically calculated values in Figure \ref{mombal_app} of the main text, labelled as the L-H scaling.
The L-H scaling captures the thrust qualitatively but not quantitatively, further indicating the need for the model derived in this study.

\bibliographystyle{jfm}
\bibliography{bibfile.bib}

\end{document}